\documentclass{aa}  
\usepackage{color}
\usepackage{rotating}
\usepackage{hyperref} 
\usepackage{tablefootnote}
\usepackage[dvipsnames]{xcolor}
\usepackage{longtable}
\usepackage{pdflscape}
\usepackage{tikz}
\def\checkmark{\tikz\fill[scale=0.4](0,.35) -- (.25,0) -- (1,.7) -- (.25,.15) -- cycle;} 
\usepackage{amsmath}

\newcommand{\struutup}{\rule{0ex}{3.2ex}}
\newcommand{\struutdown}{\rule[-2ex]{0ex}{2ex}}

\usepackage[varg]{txfonts}
\bibpunct{(}{)}{;}{a}{}{,}

\begin{document} 
   \title{KIC~8975515: a fast-rotating ($\gamma$ Dor - $\delta$ Sct) hybrid star with Rossby modes and a slower $\delta$ Sct companion in a long-period orbit }
   \authorrunning{A. Samadi-Ghadim, et~al.}
   \titlerunning{KIC~8975515: a fast-rotating hybrid star with $r$ modes and a slow-rotating $\delta$ Scuti companion }
   \author{
           A. Samadi-Ghadim\inst{1,2}
          \and
           P. Lampens\inst{3} 
          \and
           D. M. Jassur\inst{4}
           \and
           P. Jofré\inst{1}}
   \institute{Núcleo de Astronomía, Facultad de Ingeniería y Ciencias, Universidad Diego Portales, Av. Ejército Libertador 441, Santiago, Chile
   \email{anya.samadi@mail.udp.cl}
   \and
   Research Institute for Applied Physics and Astronomy, University of Tabriz, P.O.Box 51664, Tabriz, Iran
   \and
   Koninklijke Sterrenwacht van Belgi\"e, Ringlaan 3, B-1180 Brussel, Belgium
   \and
   Department of theoretical Physics and Astrophysics, Physics Faculty, University of Tabriz, P.O. Box 51664, Tabriz, Iran}
   \date{April 8, 2020; submitted }
  \abstract
  {} 
   {KIC~8975515 is a \emph{Kepler} double-lined spectroscopic binary system with hybrid pulsations. Two components have similar atmospheric properties (T$_{\rm eff}$ $\sim$ 7400~K), and one of them is a fast rotator ($v\sin i = 162$ versus 32 km/s). Our aim is to study the \emph {Kepler} light curve in great detail in order to determine the frequencies of the pulsations, to search for regular spacing patterns in the Fourier spectrum, if any, and to discuss their origin in the context of binarity and fast rotation. }
   {In this paper, we study the properties of the stellar pulsations based on a careful analysis in the low-, intermediate- and high-frequency regions of the Fourier spectrum. This is done by performing repeated frequency-search analyses with successive prewhitenings of all the significant frequencies detected in the spectrum. Moreover, we searched for regular period spacings among the $g$ modes, as well as frequency splitting among the $g$ and $p$ modes. }
   {In the low-frequency regime, five regular period spacing patterns including one series of prograde $g$ modes and four series of retrograde $r$ modes were detected. The $r$ modes are well-distributed with respect to the harmonics of the rotational frequency of the fast-rotating star $f_{\rm rot}$ = 1.647 d$^{-1}$. The dominant $g$ mode is $f_{2}$ = 2.37 d$^{-1}$. The strongest p mode, at $f_{1}$ = 13.97 d$^{-1}$, forms a singlet. In the high-frequency region, we identified two multiplets of regularly split $p$ modes with mean frequency spacings of 0.42 d$^{-1}$ and 1.65 d$^{-1}$.}
  {We detected some series of retrograde $r$ and prograde $g$ modes as well as two multiplets of $p$ modes with frequency spacings related to the stellar rotation of both components of the twin system KIC~8975515. We identified the fast-rotating component as a hybrid pulsator with $r$ modes and the slowly-rotating component as a $\delta$ Sct pulsator.}
  
\keywords{Techniques: photometric (Fourier) -
          Techniques: spectroscopic -
          (Stars:) binaries: spectroscopic -
          Stars: variables: $\delta$ Scuti -
          Stars: variables: $\gamma$ Dor - 
          Asteroseismology}
\maketitle

\section{Introduction} \label{sec:intro}
Asteroseismology is a formidable tool in astrophysics, which allows us to investigate the interior structure of stars based on phenomena observed at their surface. This knowledge is essential for a detailed understanding of both stellar structure and stellar evolution. $\gamma$ Dor/$\delta$ Scuti variables, two subgroups of classical A/F-type pulsators, are intriguing objects with regard to their stellar interior. They are located at the cross-section of the Cepheid instability strip and the main sequence. The first subgroup comprises the $\gamma$ Dor variables of spectral type A7-F5. They have masses, M, from 1.5 to 1.8 M$_{\odot}$ and temperatures, T$_{\rm eff}$, from 6700 to 7400 K \citep{Catelan2015}. Their pulsations are mainly due to low-degree ($\ell \ll 4$) high-radial order ($20 \lesssim n \lesssim 120$) gravity ($g$) modes \citep{VanReeth2016, Saio2018b} which occur in the radiative zones close to the stellar core. The fact that these modes are excited in the deep stellar interior provides us with important information about the chemical gradient of the different layers. As a result of the excited high-radial order $g$ modes \citep{Shibahashi1979,Tassoul1980}, the occurrence of deviations from a uniform period spacing reveals the chemical inhomogeneities of the near-core convective structures \citep{Miglio2008}. The typical period of their pulsation modes lies in the range from 0.3 to 3 days. Studying the pulsations of $\gamma$ Dor stars with intermediate to fast rotation provides information on their differential rotation, the angular momentum transport between the layers \citep[\emph{e.g.}][]{Ouazzani2017, Li2019a, Li2019b} and other physical processes from the different excitation layers.\\ 
\cite{VanReeth2016} detected the first observational evidence of $r$ modes (global Rossby waves) - alongside $g$ modes - for $\gamma$ Dor stars with significant rotation. Recently, \cite{Li2019b} reported the detection of both $r$ and $g$ mode period spacing patterns for 82 $\gamma$ Dor stars. In the absence of considerable stellar rotation, toroidal motions associated to $r$ modes cannot provide any compression nor expansion. Hence, the stellar atmospheres undergo neither any restoring force nor any light variations. However, in the rapidly rotating regime the toroidal motions get coupled with spheroidal motions and they present themselves as temperature perturbations. Furthermore, coupling of spheroidal motions provides excitation of the $r$ modes by the $\kappa$ mechanism \citep[\emph{e.g.} DA white dwarfs in][]{Saio1982,Berthomieu1983}. $r$ modes appear at lower frequencies than the (prograde) $g$ mode frequencies and their period spacings form a retrograde pattern \citep{VanReeth2016,Saio2018b,Li2019b}. \cite{Saio2018b} showed that $r$ modes of azimuthal order $m$ appear at frequencies lower than $m$ times the rotational frequency. \\
The second subgroup, the $\delta$ Scuti variables are intermediate-mass, pre-main sequence \citep{Zwintz2014}, main sequence \citep{Aerts2010} or post-main sequence \citep{Breger2007} stars in the classical instability strip. They are slightly hotter than $\gamma$ Dor stars with masses, M, from 1.5 to 2.5 M$_{\odot}$ and temperatures, T$_{\rm eff}$, from 6900 to 8900 K. Their pulsation periods range from 0.01 d to 0.25 d (15 min to 6 hr) \citep{Aerts2010, Catelan2015}. This group pulsates in radial and non-radial, low-degree ($\ell = 1-3$) and low-overtone ($n = 0,1,2,3,...$) $p$ modes \citep{Aerts2010, SanchezArias2017}. The $p$ modes are excited near the surface and reflect the physical properties of the stellar envelope. \\
Based on theoretical computations, \cite{Xiong2016} showed that the $\delta$ Scuti and the $\gamma$ Dor pulsators may describe a single, larger instability strip. In the region where both theoretical instability strips overlap, we expect to find hybrid behaviour \citep{Balona2011, Xiong2016}. This is clearly seen in the observational colour-magnitude diagrams, \textit{e.g.} from \emph{Kepler} \citep{Bradley2015}. Hybrid stars must have both detectable $g$ modes (in the low(er)-frequency range) and detectable $p$ modes (in the high(er)-frequency range) in their Fourier spectrum.\\
The mode driving mechanism(s) for hybrid stars ($\gamma$ Dor/$\delta$ Sct or $\delta$ Sct/$\gamma$ Dor) is (are) not well-understood. For all the classical pulsators located in the Cepheid instability strip, \emph{e.g.} the $\delta$ Scuti stars, the $\kappa$ mechanism is believed to be the major excitation mechanism \citep{Handler1999, Houdek1999, Balona2015}. Furthermore, turbulent pressure may also contribute in driving $p$ modes in $\delta$ Sct stars \citep{Houdek2000, Antoci2014}. Whereas, for $\gamma$ Dor stars, convective flux blocking (at the bottom of the convection zone) is thought to drive the pulsations \citep{Guzik2000, Dupret2004, Dupret2005}. The study by \cite{Xiong2016}, shows that for the hot $\delta$ Scuti and $\gamma$ Dor stars, the (radiative) $\kappa$ mechanism is the main driving force for both $p$ and $g$ modes, whereas, for the cool $\delta$ Sct/$\gamma$ Dor stars, it is the coupling of convection and oscillations which excites or damps the pulsations. The much larger instability strip described by \cite{Xiong2016} extends well beyond the borders of the classical instability strips of the $\gamma$ Dor and the $\delta$ Sct stars, respectively \citep[\emph{e.g.}][]{Handler1999, McNamara2000, Dupret2004}. Indeed, from an observational perspective, the width of the $\delta$ Scuti instability strip was shown to extend beyond the theoretical edges \citep{Bowman2018}. \cite{Murphy2019} derived a new empirical instability strip that is much wider (and somewhat hotter) than the theoretical ones \citep[\emph{e.g.}][]{Handler1999}.\\
The detection of pulsations in binary systems not only provides an additional tool to derive stellar parameters in an independent way but is also the perfect laboratory to check the influence of binarity (\emph{e.g.} eccentricity, tidal effects \citep{Samadi2018b,Guo2017}, mass transfer \citep{Mkrtichian2018}, chemical peculiarities \citep[][]{Kolbas2012}, presence of a third body \citep{Samadi2010,Gies2015}, etc.) on stellar structure, stellar evolution and the excited oscillation modes. Based on various surveys, the fraction of multiple systems in a population, MF, among the intermediate-mass stars is $MF^{MS}_{1.5-5M_{\odot}}\geq 50\%$ \citep{Duchene2013}. Moreover, the observed frequency of spectroscopic binaries among field intermediate-mass stars is in the range of 30-45\% \citep{Duchene2013}. Thus, we may expect that a significant fraction of the $\gamma$ Dor/$\delta$ Scuti stars resides in binary or multiple systems. Eventually, more than 2200 {\it Kepler} main-sequence A/F-type stars were studied using the pulsation timing method in search of binarity across a narrow interval in (log) period allowed by the method \citep{Murphy2018}. These authors detected a binary fraction of 13.9$\pm$2.1\% among the studied sample. In comparison, \cite{Lampens2018} reported an extensive multi-epoch spectroscopic survey of 50 \textit{Kepler} hybrid pulsators for which they derived a multiplicity fraction of at least 27\%.\\
Stellar pulsation studies exploiting ever more precise photometric data from space missions such as \emph{CoRoT} \citep{Baglin2006}, \emph{Kepler} \citep{Koch2010} (\emph{K2}), \emph{TESS} \citep{Ricker2015} have become very fruitful as these missions provide light curves of unprecedented accuracy with errors of the order of few $\mu$mag (\emph{i.e.} few parts per million). These space missions allowed the detection of several hundreds of hybrid $\gamma$ Dor/$\delta$ Sct pulsators exhibiting both types of modes simultaneously, be it with very low amplitudes. For a detailed historical review of the discoveries, we refer to \cite{Qian2019}. We briefly review a few extensive studies concerning systems with hybrid pulsations here. Examples of such studies are: KIC 4544587, an eccentric eclipsing binary system with an orbital period P$_{\rm orb}$ = 2.19 days \citep{Hambleton2013}; KIC 10080943, a double-lined spectroscopic binary with P$_{\rm orb}$ = 15.34 days \citep{Schmid2015, Keen2015}; KIC~9592855, a post-mass-transfer eclipsing binary with P$_{\rm orb}$ = 1.2 days \citep{Guo2017}; KIC~4150611, a quintuplet system where the primary is a triplet \citep{Helminiak2017} and KIC 6048106, an Algol-type eclipsing binary with P$_{\rm orb}$ = 1.56 days \citep{Lee2016, Samadi2018a, Samadi2018b}. In addition, \cite{Derekas2019} discovered the double-lined spectroscopic binary KIC~5709664 (using the phase modulation (PM) method \citep{Murphy2014} and fitting of radial velocity data) with both $r$- and $p$-modes. \\
In this paper, we focus our attention on a \emph{Kepler} (candidate) hybrid pulsating star recently discovered as a double-lined,  long-period, high mass-ratio, spectroscopic binary system (SB2), KIC~8975515. Our goal is to detect the pulsation frequencies and to characterise the pulsations using the photometric {\it Kepler} data. The properties of the \emph{Kepler} light curves are described in Sec.~\ref{sec:photom_obs}. We describe our methodology for the pulsation study in Sec.~\ref{sec:pulsation_method}. The details and results of the frequency analyses  for both the high- and the low-frequency regions are presented in Sec.~\ref{sec:fourier}. We discuss all the results brought together in Sec.~\ref{sec:discussion}. Finally, in Sec.~\ref{sec:con}, we present a summary and our main conclusions.
\section{Photometric observations from the \emph{Kepler} mission} \label{sec:photom_obs}
\begin{figure}
 \centering
	\includegraphics[width=\columnwidth]{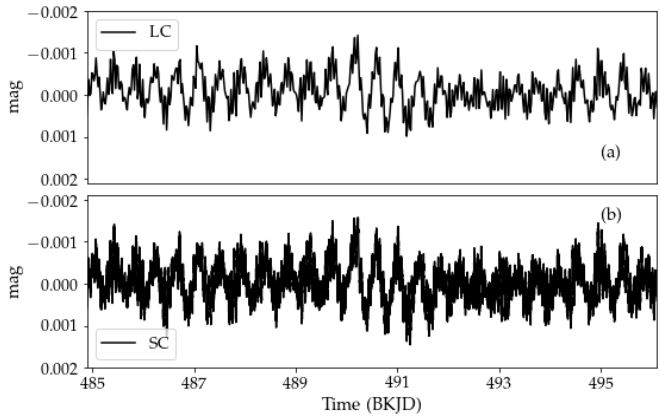}
	\caption{Close-up view of the light variations of KIC~8975515. The vertical axis displays the relative magnitude (mag) while the horizontal axis displays the Barycentric \emph{Kepler} Julian Date (BKJD = BJD - 2454833.0) in days. (a) An 11-day interval of the Long-Cadence (LC) light curve from Q6. (b) The same time interval of the Short-Cadence (SC) light curve.}
	\label{fig:lc_zoom}
\end{figure}
\begin{table}
 \centering
	\caption{Information on KIC~8975515 from surveys and data bases.}
	\label{tab:photom_data}
	\begin{tabular}{lccc}
	\hline\hline
	{\textbf Survey}     & {\textbf Target ID} \\
	\hline
	\emph{Kepler}     & KIC~8975515 \\
	2MASS      & J19534557+4513092\\
	TYC        & 3558-2169-1\\
	Gaia       & 2079403021883452288 \\
	\hline\hline
	& KIC$^{(1)}$\\
	\hline\hline
	{\textbf Parameter}         & {\textbf Value}  \\
	\hline
	RA                 & +19$^{h}$:53$^{m}$:45.5766$^{s}$ \struutup\\
	Dec                & +45$\degr$:13$\arcmin $:9.274$\arcsec$\\
	LC                 & Q0-Q17 \\
	SC                 & Q2.3, Q5.1,5.2,5.3 \\
    BJD$_\mathrm{0}$ (LC)(d)  &  2454953.5385\\
    Kmag (mag)        & 9.515  \\
	T$_\mathrm{eff}$ (K)  & 7176  \\
	$\log{g}$ (cgs)       & 3.896  \\
    R (R$_{\odot}$)  & 2.197\struutdown\\               
    \hline\hline
    & GAIA DR2$^{(2)}$\\
    \hline\hline
     $\varpi$ (mas)          & 2.506$\pm$0.0387 \struutup\\
     m$_{\rm G}^{(3)}$ (mag) & 9.447 \\
     T$_\mathrm{eff}$ (K)    & 7157 \struutdown\\
      \hline\hline
     &Spectroscopy$^{(4)}$ \\
     \hline\hline
     	{\textbf Parameter} & {\textbf Primary} & {\textbf Secondary}\\
	\hline
    $v \sin{i}$ (km/s)  & 162$\pm$2      & 32$\pm$1 \struutup\\
    T$_{\rm eff}$ (K)   & 7440$\pm$250$^{(5)}$ & 7380$\pm$250 \struutdown\\
    \hline
	{\textbf Parameter} & {\textbf Value$^{(6)}$} & {\textbf Uncertainty}\\
	\hline
	P$_{\rm orb}$ (d) & 1603 & 9 \struutup\\
	e                 & 0.408 & 0.015\\
    q                 & 0.83  & 0.05 \struutdown\\
    \hline
    \end{tabular}
    \tablebib{(1) from Kepler Input Catalogue (based on 5-band photometry (plus the J, H, \& K bands from the 2MASS survey) considering that the object is a single star); (2)~\cite{Gaia2018a,Gaia2018b}; (3) mean G-band magnitude; (4) \cite{Lampens2018}; (5) The uncertainty was estimated from a partial comparison of effective temperatures measured with those from the literature (Lampens et al. 2018, Fig. 7); (6) Lampens et al. (2020) \emph{in prep.} (priv. communic.)}
\end{table}
KIC~8975515, a bright \emph{Kepler} object with K$_{\rm p}$ = 9.515 mag, was observed during the quarters Q0-Q17 in Long Cadence mode (LC with a sampling of 29.42 min). There are four months of observations in Short Cadence (SC) mode available with a sampling of 58.85 s from the quarters Q2.3, Q5.1, Q5.2 and Q5.3. Altogether, we have 1470.46 days ($f_{\rm res}  = 0.00068 $ d$^{-1}$)\footnote{The frequency resolution equals (time span of the observations)$^{-1}$} of LC and 121.79 d of SC ($f_{\rm res} = 0.00821 $ d$^{-1}$) data available. For the light curves, we took the \textit{Kepler} original flux and its error for all quarters from the Kepler Asteroseismic Science Operations Centre \href{http://kasoc.phys.au.dk/}{(KASOC)}. We first converted the flux to magnitude and then applied a polynomial fit to each quarter. The fit was then subtracted from the original data in order to smooth the trends in each quarter. We finally concatenated the light curves of all quarters and removed the outliers manually. The 'detrended' light curve in LC mode was used for this pulsation study.\\
A close-up view of the light variations of KIC~8975515 is shown in Fig. \ref{fig:lc_zoom} for both LC and SC sampling modes. The relative magnitude varies with a semi-amplitude of 1 mmag on average.\\
KIC~8975515 was first studied and reported as an A/F-type hybrid star by \cite{Uytterhoeven2011}. We list the information available from \emph{Kepler} Input Catalogue in Table~\ref{tab:photom_data} (derived with the assumption of single star). Moreover, \emph{Gaia} DR2 \citep{Gaia2018a, Gaia2018b} measured a parallax of $\varpi = 2.506\pm0.0387$ mas for this target. The projected rotational velocities are those from \cite{Lampens2018}. The orbital parameters such as P$_{\rm orb}$, the eccentricity, e, and the mass ratio, q, (the ratio of a${_A}\sin i $ (Table~11)  and a${_2}\sin i $ (Table~8) reported by \cite{Lampens2018} shows that q $\sim$ 0.8) indicate that the system consists of two stars of almost similar mass in a long-period, eccentric orbit. The updated values in Table~\ref{tab:photom_data} are from Lampens et al.\, (priv. comm.). 
\section{Methodology of the pulsation study}\label{sec:pulsation_method}
Our approach is based on the Lomb-Scargle periodogram \citep{Lomb1976, Scargle1982}. We calculated the periodogram of the observed light curves up to the Nyquist frequency (both LC and SC samplings with $f_{\rm NY_{\rm LC}} = 24.65$ d$^{-1}$ and $f_{\rm NY_{\rm SC}} = 734.07$ d$^{-1}$). The Signal-to-Noise Ratio (SNR) of each frequency was calculated in a box-size of 2 d$^{-1}$. The prewhitening method is from \cite{Vanicek1971}. We consider a frequency significant only if SNR$ \geq $4, based on \cite{Breger1993}'s criterion. For detailed information on the frequency analysis and the prewhitening methods, along with the error determination see \cite{Samadi2018b}. We consider two frequencies as resolved if their difference is larger than the resolution frequency, $f_{\rm res}$ = 1/T, which is 0.00068 d$^{-1}$ for the LC light curve. This procedure provided us with a list of prewhitened frequencies of acceptable SNR and amplitude. We will refer to them as 'significant' frequencies and we will check the possible origin (whether they can be due to pulsations or caused by some other mechanism). During the frequency analysis we also checked whether any significant frequency is a linear combination of the previous frequencies of the highest SNR and larger amplitudes (i.e. having larger than mean SNR and mean amplitude of the previously prewhitened $g$ modes).   \\
According to asymptotic theory of stellar pulsations \citep{Shibahashi1979,Tassoul1980} regular period spacing may occur for the high-radial order $g$ modes, $n\gg l$, with $n$ and $\ell$ the radial order and the degree of the mode, respectively. In contrast, for the $p$ modes, regular splitting may occur because of the rotation. For genuine hybrid stars, we can expect to see these regularities in the period and frequency domains for the $g$ modes and in the frequency domain for the $p$ modes. The regular period spacings, for the $g$ and $p$ modes, can be affected by rotation or chemical inhomogeneties \citep{Bouabid2013, Li2019b}.
\section{Frequency analysis}\label{sec:fourier}
\begin{figure*}
 \centering
	\includegraphics[width=17cm]{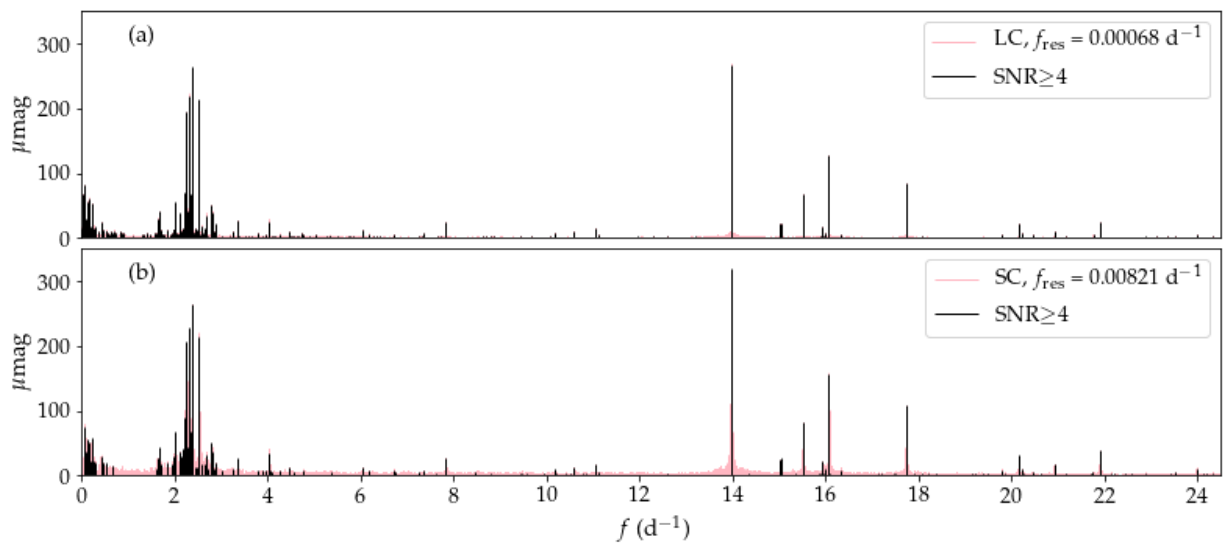}
	\caption{Fourier Spectrum of KIC 8975515 (pink). (a) Frequency spectrum associated to LC observations (pink) and the significant frequencies with SNR $\geqslant$ 4.0 in a box size of 2d$^{-1}$ (black). (b) Same as (a) but for the SC observations (referring to the quarters Q2.3 and Q5.1, 5.2, 5.3). There is no significant frequency between 25 and 50 d$^{-1}$ and we clipped it off from the illustration.}
	\label{fig:fourier_spectrum_general}
\end{figure*}
Fig.\ref{fig:fourier_spectrum_general} shows the Fourier spectrum both for the LC and SC \textit{Kepler} light curves. The full Fourier spectrum is shown in pink while the significant frequencies are presented in black. In the case of SC observations, the illustration (panel (b) in Fig.~\ref{fig:fourier_spectrum_general}) shows a close-up view of the frequency range (0-25~d$^{-1}$) where some significant frequencies show up. Clearly, both LC and SC frequency spectra present almost the same features. For the current study we used the LC light curve with the longest time base. The list of detected significant frequencies is presented in Table~\ref{tab:detec_freq}. It is sorted according to decreasing amplitude. In total we resolved 331 frequencies.\\
Interestingly, we can see that the dominant modes pertain to both the low and the high-frequency regions. Indeed, the dominant frequency in the high-frequency region, $f_{1} = 13.97236\pm 0.00001$~d$^{-1}$, has an amplitude of A$_{1} = 268\pm5~\mu$mag, while the most significant frequency in the $g$ mode region, $f_{2} = 2.37418\pm0.00001$~d$^{-1}$, occurs with very similar amplitude, A$_{2} = 265\pm5~\mu$mag.\\
Next we searched for the frequencies that are combinations of the most significant frequencies, \emph{i.e.} 'parent' frequencies. The parent frequencies were chosen among low-frequency modes up to 3.3 d$^{-1}$ and have both amplitude and SNR higher than the corresponding mean values (\emph{i.e.} A$_{\rm mean} = 13~\mu$mag and SNR$_{\rm mean} = 14.9$) of the detected frequencies. Accordingly, we selected 24 frequencies as parent frequencies. Furthermore, we searched for the combination of 13.97 d$^{-1}$ with any low-frequency mode to reveal the high frequencies which might appear as coupling of $f_{1}$ and $g$ modes. This can help to detect low- and high-frequency modes originating from the same star. Generally, a combination frequency will have its amplitude smaller than that of both parent frequencies and the difference between this linear combination (of the parent frequencies) and the candidate combination frequency should match within a tolerance of the resolution frequency \citep[\textit{e.g.}][]{Zhang2018}. The results are reported in Table~\ref{tab:detec_freq} and we indicate the parent frequencies with a check-mark under the column 'P'. The subscript '*' means that the detected frequency is not a unique combination of parent frequencies. Moreover, the frequencies that are a possible harmonic of any parent frequency are also reported in Table~~\ref{tab:detec_freq}.  
\subsection{The high-frequency region}\label{sec:pmodes}
\begin{figure*}
\centering
	\includegraphics[width=17cm]{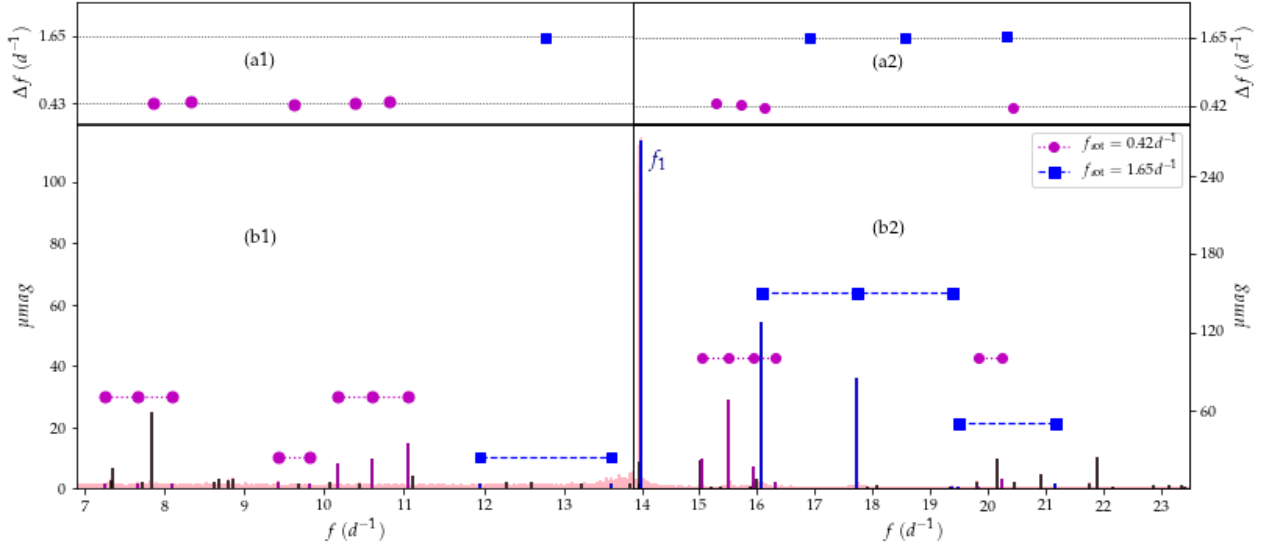}
	\caption{Regular frequency spacing for $p$ modes. (a1,a2) Frequency splittings and their deviation from mean frequency splitting (Table~\ref{tab:pmode_splt}). (b1,b2) Associated frequencies of the detected multiplets of significant frequencies (black) and the full frequency spectrum (pink). }
	\label{fig:pmode_splt}
\end{figure*}
\begin{table*}
 \centering
	\caption{$p$ modes with (quasi-)regular frequency splitting in the Fourier spectrum of KIC 8975515 (Fig.~\ref{fig:pmode_splt}). \textit{SR/FR}: detected $p$ modes are most probably associated to Slowly Rotating/Fast Rotating star. $\epsilon_{\Delta f}$ is $\sqrt{\epsilon_{f_1}^2+\epsilon_{f_2}^2}$. $\sigma_{\Delta}$ is the standard deviation.}
	\label{tab:pmode_splt}
	\begin{tabular}{lccccccc}
	\hline
	\hline
	$f_\mathrm{i}$  & $f\pm\epsilon_{f}$& $A$      & $\Pi$ & $\Delta f\pm\epsilon_\mathrm{\Delta f}$&  SNR & color & comment\struutup\\
	                & d$^{-1}$          & $\mu$mag & d     &  d$^{-1}$                              &      &   Fig.~\ref{fig:pmode_splt} (b)     & \struutdown      \\
	\hline
    $f_\mathrm{301}$ &  7.24786$\pm$0.00026   &  1.63   &  0.13797  &                        & 6    & \textcolor{Magenta}{$\bullet$} & SR\\
    $f_\mathrm{299}$ &  7.65563$\pm$0.00026   &  1.75   &  0.13062  & 0.4078$\pm$0.0004      & 7    & \textcolor{Magenta}{$\bullet$} & SR\\
    $f_\mathrm{304}$ &  8.10046$\pm$0.00027   &  1.56   &  0.12345  & 0.4448$\pm$0.0004      & 6    & \textcolor{Magenta}{$\bullet$} & SR\\
    $f_\mathrm{278}$ &  9.42582$\pm$0.00029   &  2.23   &  0.10609  &                        & 11   & \textcolor{Magenta}{$\bullet$} & SR\\
    $f_\mathrm{308}$ &  9.81926$\pm$0.00030   &  1.51   &  0.10184  & 0.3934$\pm$0.0004      & 7    & \textcolor{Magenta}{$\bullet$} & SR\\
    $f_\mathrm{116}$ &  10.16604$\pm$0.00015  &  8.08   &  0.09837  &                        & 31   & \textcolor{Magenta}{$\bullet$}& SR\\
    $f_\mathrm{94}$  &  10.59499$\pm$0.00013  &  9.72   &  0.09438  & 0.4289$\pm$0.0002      & 30   & \textcolor{Magenta}{$\bullet$} & SR\\
    $f_\mathrm{51}$  &  11.04825$\pm$0.00009  &  14.65  &  0.09051  & 0.4532$\pm$0.0001      & 30   & \textcolor{Magenta}{$\bullet$} & SR\\
    $f_\mathrm{32}$  &  15.05250$\pm$0.00006  &  22.82  &  0.06643  &                        & 30   & \textcolor{Magenta}{$\bullet$} & SR\\
    $f_\mathrm{9}$   &  15.51470$\pm$0.00003  &  68.71  &  0.06446  & 0.4621(9)$\pm$0.00006  & 77   & \textcolor{Magenta}{$\bullet$} & SR\\
    $f_\mathrm{45}$  &  15.94343$\pm$0.00008  &  16.27  &  0.06272  & 0.4287(3)$\pm$0.00009  & 30   & \textcolor{Magenta}{$\bullet$} & SR\\
    $f_\mathrm{193}$ &  16.32732$\pm$0.00022  &  5.00   &  0.06125  & 0.3839$\pm$0.0002      & 27   & \textcolor{Magenta}{$\bullet$} & SR\\
    $f_\mathrm{326}$ &  19.35654$\pm$0.00036  &  1.16   &  0.05166  &                        & 6    & \textcolor{Magenta}{$\bullet$} & SR\\
    $f_\mathrm{184}$ &  19.80269$\pm$0.00022  &  5.28   &  0.05050  & 0.4461$\pm$0.0004      & 31   & \textcolor{Magenta}{$\bullet$} & SR\\
    $f_\mathrm{127}$ &  20.23788$\pm$0.00017  &  7.31   &  0.04941  & 0.4352$\pm$0.0003      & 30   & \textcolor{Magenta}{$\bullet$} & SR\\[5pt]
                     &                        &          &  $\Delta f_{\rm mean}\pm \sigma_{\Delta}$ & 0.419$\pm$0.020 &  & \\  [8pt] 
    $f_\mathrm{303}$ &  11.95624$\pm$0.00030  &  1.60    &  0.08364 &                        &  7   &\textcolor{Blue}{$\blacksquare$} & FR \\
    $f_\mathrm{320}$ &  13.58882$\pm$0.00036  &  1.27    &  0.07359 & 1.6326$\pm$0.0005      &  7   &\textcolor{Blue}{$\blacksquare$} & FR \\
    $f_\mathrm{6 }$  &  16.07633$\pm$0.00002  &  128.14  &  0.06220 &                        &  76  &\textcolor{Blue}{$\blacksquare$} & FR \\
    $f_\mathrm{7 }$  &  17.73452$\pm$0.00002  &  84.42   &  0.05639 & 1.6581(9)$\pm$0.00003  &  77  &\textcolor{Blue}{$\blacksquare$} & FR \\
    $f_\mathrm{305}$ &  19.38475$\pm$0.00039  &  1.55    &  0.05159 & 1.6502$\pm$0.0004      &  9   &\textcolor{Blue}{$\blacksquare$} & FR \\
    $f_\mathrm{311}$ &  19.48475$\pm$0.00037  &  1.42    &  0.05132 &                        &  8   &\textcolor{Blue}{$\blacksquare$} & FR \\
    $f_\mathrm{249}$ &  21.16071$\pm$0.00027  &  3.00    &  0.04726 & 1.6759$\pm$0.0005      &  19  &\textcolor{Blue}{$\blacksquare$} & FR \\[5pt]	     
	               &                       &        &  $\Delta f_{\rm mean}\pm \sigma_{\Delta}$ & 1.654$\pm$0.018 &  & \\[8pt]                
	\hline
	\end{tabular}
\end{table*}
\begin{figure}
\centering
	\includegraphics[width=\columnwidth]{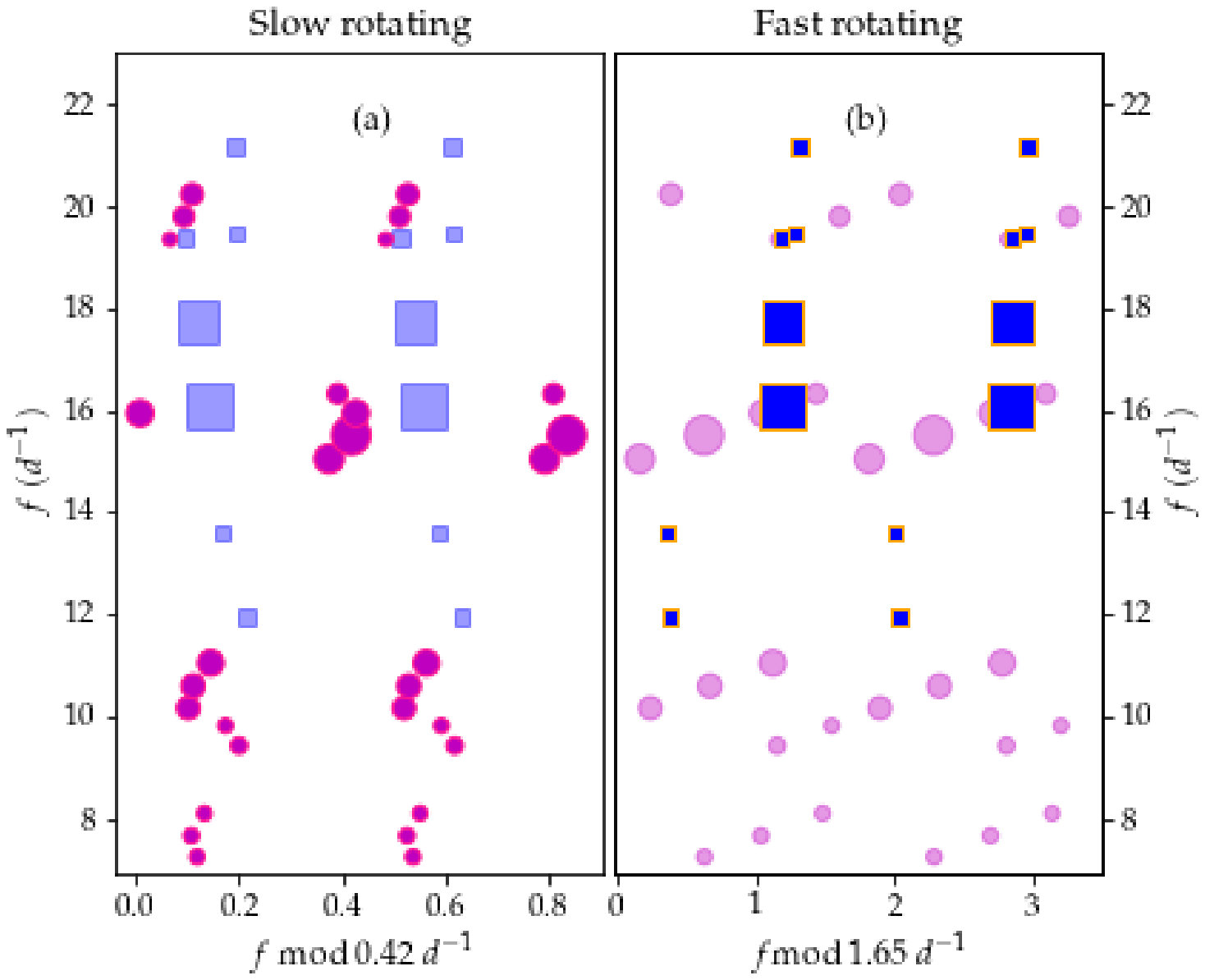}
	\caption{\'{E}chelle diagram of $p$ modes with (quasi-)regular frequency spacing from the Fourier spectrum of KIC~8975515 (Table~\ref{tab:pmode_splt}). Symbol size refers to the amplitude of the $p$ modes. Colour coding and symbols are similar to Fig.~\ref{fig:pmode_splt}. (a) frequency modulo the mean rotational frequency spacing value associated to the slowly-rotating star, 0.419 d$^{-1}$ (magenta circles) plotted twice. (b) frequency modulo the mean rotational frequency spacing value associated to the fast-rotating star, 1.654 d$^{-1}$ (blue squares) plotted twice. }
	\label{fig:pmode_freqechelle}
\end{figure}
A quick look at Fig.~\ref{fig:fourier_spectrum_general} and a close-up view of the high-frequency section of Fig~\ref{fig:pmode_splt} (b) shows that the $p$ modes are very sparsely distributed and extend to 22 d$^{-1}$, unlike the distribution of the $g$ modes (\ref{sec:gmodes_pspacing}). The significant frequencies have amplitudes ranging from A = 1 to 268 $\mu$mag.\\
Because of non-linear effects, the combination of the $g$ modes, $f_{g}$, and the dominant $p$ mode ($f_{p_{max}}$) may generate some coupled frequency peaks in the form of $f_{p_{max}} \pm f_{g}$ \citep{Kurtz2014}. We found $f_{303}$ (11.96 d$^{-1}$) to be a combination of $f_{1}$ and $f_{16}$ (2.79 d$^{-1}$) that is a prograde $g$ mode. On the other hand $f_{285}$ (12.29 d$^{-1}$) is a combination of $f_{1}$ and $f_{18}$ which is a retrograde mode. Similarly $f_{94}$ (10.59 d$^{-1}$) is a combination of $f_{1}$ with another retrograde mode ($f_{35}$).  We refer to these frequencies with 'MC' in Table~\ref{tab:detec_freq}. We ran a Monte Carlo simulation and found that each combination has about 2\% probability of occurring by random chance.\\
We detected two different types of multiplets: doublets and triplets with the (semi-)regular frequency spacing of $\Delta f_{\rm mean_{\bullet}} = 0.419\pm0.020$ d$^{-1}$ (magenta in Fig.~\ref{fig:pmode_splt}) and $\Delta f_{\rm mean_{\blacksquare}} = 1.654\pm0.018$ d$^{-1}$ (blue in Fig.~\ref{fig:pmode_splt}), located on either side of $f_{1}$, though not involving $f_{1}$. The quoted error is the standard deviation. Table~\ref{tab:pmode_splt} lists the associated frequencies in d$^{-1}$ and  the  $\Delta f$ value between each couple of them where $\epsilon_{\Delta f}$ equals $\sqrt{\epsilon_{f_1}^2+\epsilon_{f_2}^2}$. Fig.~\ref{fig:pmode_splt}, presents the associated frequencies (lower panel) and their $\Delta f$ values (upper panel) distinguishable by two different colours and symbols, respectively. We detected one frequency in the $g$ mode region $f_{200}$ (1.6467$\pm$0.0001 d$^{-1}$) which is equal to the larger frequency splitting within the errors ($1.654\pm0.018$ d$^{-1}$). However, there are other groups of frequencies in the interval (6.72 to 8.63 d$^{-1}$) that are either a combination of prograde $g$ modes ($f_{156}, f_{262}$) or a combination of prograde $g$ and retrograde $r$ modes ($f_{145}, f_{28}$ and $f_{281}$). We suggest that the larger frequency splitting is related to the fast rotating companion and the smaller is related to the slowly rotating one. We will present arguments in favor in Sec.~\ref{sec:discussion}.\\
We present the \'{e}chelle diagram (Fig.~\ref{fig:pmode_freqechelle}) for all the significant high frequencies, using the large mean frequency splitting from Table~\ref{tab:pmode_splt}. Panel (a) in Fig.~\ref{fig:pmode_freqechelle} shows the frequency modulo the mean frequency splitting associated to the slowly rotating star $f_{\rm rot} = 0.419$ d$^{-1}$. The symbol size is associated to the mode amplitudes and the colour coding is similar to Fig.~\ref{fig:pmode_splt}. We highlighted these frequencies using magenta circles. Similarly in panel (b) we show the modes with detected regular splittings versus modulo $f_{\rm rot} =1.654$ d$^{-1}$ which we associated to the fast rotating star. We highlighted the modes with this average splitting using blue squares.\\
\subsection{The low-frequency region}\label{sec:gmodes_pspacing}
\begin{figure*}
	\includegraphics[width=17cm]{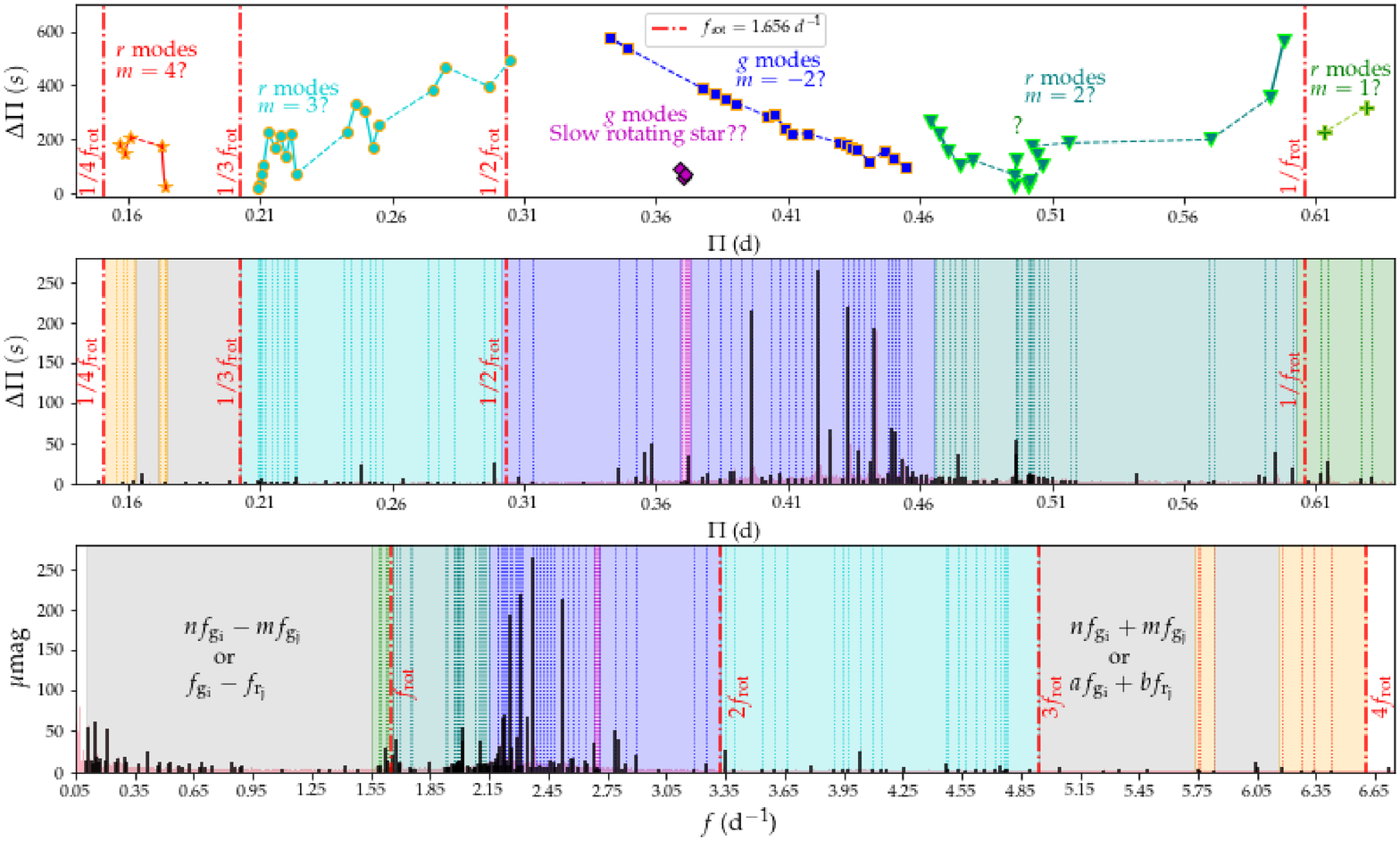}
	\centering
	\caption{(a) Period spacing patterns for $r$ and $g$ modes on the period spectrum of KIC~8975515 (LC). (b) The associated frequencies for the detected $g$/$r$ modes in (a) (Tables~\ref{tab:lowfreq_gmode_pspacing} \& \ref{tab:lowfreq_rmode_pspacing}) showed with pointed-lines in the same colour in panel (a). The frequencies in black are the significant modes and the full periodogram of KIC~8975515 is plotted in light pink.}
	\label{fig:lowfreq_pspacing}
\end{figure*}
\begin{table*}
 \centering
	\caption{Detected $g$ modes in the Fourier spectrum of KIC~8975515 (Long Cadence observations). \textit{FR} suggests that the suggested origin of the modes is the fast-rotating star ($v\sin i =162 $ km/s).  }
	\label{tab:lowfreq_gmode_pspacing}
	\begin{tabular}{lcccccccc}
	\hline
	\hline
	   $f_\mathrm{i}$  & $f\pm\epsilon_{f}$ & A         & $\Pi$   & $\Delta \Pi \pm\epsilon_{\rm \Delta \Pi}$ & SNR            & colour & comment\struutup\\
	                   & d$^{-1}$           & $\mu$mag  & d       & s                                         &                & Fig. \ref{fig:lowfreq_pspacing}&\struutdown\\
\hline
 &\\
 $f_{\mathrm 37}$  &  2.89346$\pm$0.00006  &  20.83  &  0.34561  &             & 31  & \textcolor{Blue}{$\blacksquare$}& FR; $m=-2$?\\      
 $f_{\mathrm 83}$  &  2.83861$\pm$0.00008  &  9.61   &  0.35229  & 577$\pm$312 & 15  & \textcolor{Blue}{$\blacksquare$}& FR; $m=-2$?\\      
 $f_{\mathrm 95}$  &  2.78938$\pm$0.00003  &  51.37  &  0.35850  & 537$\pm$272 & 30  & \textcolor{Blue}{$\blacksquare$}& FR; $m=-2$?\\      
 $f_{\mathrm 54}$  &  2.63465$\pm$0.00006  &  14.21  &  0.37956  &             & 19  & \textcolor{Blue}{$\blacksquare$}& FR; $m=-2$?\\      
 $f_{\mathrm 147}$ &  2.60372$\pm$0.00009  &  6.69   &  0.38407  & 390$\pm$124 & 10  & \textcolor{Blue}{$\blacksquare$}& FR; $m=-2$?\\      
 $f_{\mathrm 40}$  &  2.57491$\pm$0.00005  &  16.79  &  0.38836  & 371$\pm$106 & 22  & \textcolor{Blue}{$\blacksquare$}& FR; $m=-2$?\\      
 $f_{\mathrm 157}$ &  2.54831$\pm$0.00010  &  6.18   &  0.39242  & 350$\pm$85  & 9   & \textcolor{Blue}{$\blacksquare$}& FR; $m=-2$?\\      
 $f_{\mathrm 4}$   &  2.52345$\pm$0.00001  &  213.96 &  0.39628  & 334$\pm$69  & 70  & \textcolor{Blue}{$\blacksquare$}& FR; $m=-2$?\\
 $f_{\mathrm 72}$  &  2.47839$\pm$0.00007  &  11.37  &  0.40349  &             & 13  & \textcolor{Blue}{$\blacksquare$}& FR; $m=-2$? \\
 $f_{\mathrm 49}$  &  2.45825$\pm$0.00005  &  14.90  &  0.40679  & 286$\pm$22  & 15  & \textcolor{Blue}{$\blacksquare$}& FR; $m=-2$? \\  
 $f_{\mathrm 146}$ &  2.43813$\pm$0.00009  &  6.69   &  0.41015  & 290$\pm$26  & 9   & \textcolor{Blue}{$\blacksquare$}& FR; $m=-2$? \\
 $f_{\mathrm 152}$ &  2.42171$\pm$0.00009  &  6.43   &  0.41293  & 240$\pm$24  & 9   & \textcolor{Blue}{$\blacksquare$}& FR; $m=-2$? \\
 $f_{\mathrm 143}$ &  2.40690$\pm$0.00009  &  6.77   &  0.41547  & 220$\pm$44  & 9   & \textcolor{Blue}{$\blacksquare$}& FR; $m=-2$? \\
 $f_{\mathrm 182}$ &  2.38873$\pm$0.00010  &  5.34   &  0.41863  &             & 8   & \textcolor{Blue}{$\blacksquare$}& FR; $m=-2$?\\      
 $f_{\mathrm 2}$   &  2.37418$\pm$0.00001  &  264.99 &  0.42120  & 222$\pm$44  & 77  & \textcolor{Blue}{$\blacksquare$}& FR; $m=-2$?\\      
 $f_{\mathrm 140}$ &  2.32178$\pm$0.00008  &  6.93   &  0.43070  &             & 9   & \textcolor{Blue}{$\blacksquare$}& FR; $m=-2$?\\      
 $f_{\mathrm 3}$   &  2.31022$\pm$0.00001  &  219.12 &  0.43286  & 186$\pm$79  & 79  & \textcolor{Blue}{$\blacksquare$}& FR; $m=-2$?\\      
 $f_{\mathrm 129}$ &  2.29919$\pm$0.00008  &  7.23   &  0.43494  & 179$\pm$86  & 9   & \textcolor{Blue}{$\blacksquare$}& FR; $m=-2$?\\      
 $f_{\mathrm 17}$  &  2.28902$\pm$0.00003  &  41.86  &  0.43687  & 167$\pm$98  & 31  & \textcolor{Blue}{$\blacksquare$}& FR; $m=-2$?\\      
 $f_{\mathrm 219}$ &  2.27908$\pm$0.00010  &  4.30   &  0.43877  & 164$\pm$101 & 6   & \textcolor{Blue}{$\blacksquare$}& FR; $m=-2$?\\      
 $f_{\mathrm 25}$  &  2.26457$\pm$0.00004  &  28.55  &  0.44158  &             & 27  & \textcolor{Blue}{$\blacksquare$}& FR; $m=-2$?\\      
 $f_{\mathrm 5}$   &  2.25761$\pm$0.00001  &  194.15 &  0.44295  & 118$\pm$147 & 80  & \textcolor{Blue}{$\blacksquare$}& FR; $m=-2$?\\
 $f_{\mathrm 50}$  &  2.23396$\pm$0.00005  &  14.84  &  0.44763  &             & 15  & \textcolor{Blue}{$\blacksquare$}& FR; $m=-2$?\\      
 $f_{\mathrm 8}$   &  2.22490$\pm$0.00002  &  70.43  &  0.44946  & 157$\pm$108 & 30  & \textcolor{Blue}{$\blacksquare$}& FR; $m=-2$?\\      
 $f_{\mathrm 11}$  &  2.21884$\pm$0.00002  &  65.65  &  0.45069  &             & 30  &\textcolor{Blue}{$\blacksquare$}& FR; $m=-2$?\\      
 $f_{\mathrm 107}$ &  2.21165$\pm$0.00007  &  8.82   &  0.45215  & 127$\pm$137 & 11  & \textcolor{Blue}{$\blacksquare$}& FR; $m=-2$?\\      
 $f_{\mathrm 31}$  &  2.19728$\pm$0.00004  &  23.44  &  0.45511  &             & 22  & \textcolor{Blue}{$\blacksquare$}& FR; $m=-2$?\\      
 $f_{\mathrm 109}$ &  2.19179$\pm$0.00007  &  8.77   &  0.45625  & 97$\pm$167  & 10  & \textcolor{Blue}{$\blacksquare$}& FR; $m=-2$?\\[5pt]
                   &                       &         &  $\Delta \Pi_{\rm mean}\pm \sigma_{\Delta}$ & 264$\pm$132   &    \\[8pt]
 \hline
 &\\
 $f_{\mathrm 237}$ &  2.70471$\pm$0.00012  &  3.63   &  0.36972  &           & 7   & \textcolor{Magenta}{$\blacklozenge$}   &  ? \\    
 $f_{\mathrm 204}$ &  2.69696$\pm$0.00012  &  4.72   &  0.37079  & 92$\pm$18 & 9   &\textcolor{Magenta}{$\blacklozenge$}    &  ?  \\   
 $f_{\mathrm 228}$ &  2.69193$\pm$0.00012  &  3.91   &  0.37148  & 60$\pm$13 & 7   &\textcolor{Magenta}{$\blacklozenge$}    &  ?  \\   
 $f_{\mathrm 22}$  &  2.68620$\pm$0.00004  &  34.90  &  0.37227  & 68$\pm$5  & 30  &\textcolor{Magenta}{$\blacklozenge$}    &  ?  \\ [5pt]
                  &                       &         &  $\Delta \Pi_{\rm mean}\pm \sigma_{\Delta}$ & 73$\pm$13   &    \\[5pt]
\hline
    \end{tabular}
\end{table*}
We carefully investigated the period spectrum in search of any regular period spacing pattern. The upper panel in Fig.~\ref{fig:lowfreq_pspacing} shows different patterns of period spacing $\Delta \Pi$ that are either prograde (blue squares) or retrograde (green triangles and cyan circles). Their associated periods and frequencies with similar colours are illustrated in the middle and lower panels, respectively. The period spacings and their deviation from the mean value, $\epsilon_{\Delta \Pi}$, are reported in Tables~\ref{tab:lowfreq_gmode_pspacing} \& \ref{tab:lowfreq_rmode_pspacing}. For the prograde pattern in the frequency interval (2.19 to 2.90) d$^{-1}$, $\Delta \Pi$ decreases from 577 to 97 s with increasing period. Its mean period spacing equals $\Delta \Pi_{\rm mean_{\blacksquare}} = 264\pm132$ s (the error is the standard deviation). There are several frequencies that are a combination of parent prograde $g$ modes but do fit very well in the pattern (\emph{e.g.} in the interval (2.40 to 2.47) d$^{-1}$). In addition, in the frequency interval (2.68 to 2.70) d$^{-1}$ we detected a group of frequencies of much lower $\Delta \Pi_{\rm mean_{\blacklozenge}} = 73\pm13$ s that are well separated from the pre-cited prograde pattern. We show them with magenta diamonds in Fig.~\ref{fig:lowfreq_pspacing} (for the frequencies see Table~\ref{tab:lowfreq_gmode_pspacing}). The retrograde modes in the frequency interval (3.24 to 4.79) d$^{-1}$ have $\Delta \Pi_{\rm mean_{\bullet}} = 226\pm137$ s (cyan circles in Fig.~\ref{fig:lowfreq_pspacing}). However, some of these modes (in Table~\ref{tab:lowfreq_rmode_pspacing}) are also combinations of prograde $g$ modes \citep{Saio2018b}. These combinations are listed in Table~\ref{tab:detec_freq}. For these modes, $\Delta \Pi$ increases from 20 to 491 s with increasing period. The second group of retrograde modes was detected in the range (1.66 to 2.15) d$^{-1}$. They are illustrated by the green triangles in Fig.~\ref{fig:lowfreq_pspacing}. Though the lower periods show a decreasing period spacing, we consider them to be part of the same group as discussed in Sec~\ref{sec:discussion}. The list of frequencies and their period spacing values are reported in Table~\ref{tab:lowfreq_rmode_pspacing}. $\Delta \Pi$ in this pattern first decreases from 268 to 123 s and then increases from 28 to 565 s and its overall average is $\Delta \Pi_{\rm mean_{\blacktriangle}} = 162\pm130$ s. Similar to previous group of retrograde modes some of these modes are also a combination of prograde $g$ modes (Table~\ref{tab:detec_freq}). Furthermore, we detected two more period spacing patterns (retrograde see Sec.~\ref{sec:discussion}) with a smaller number of detected modes than other detected modes. These modes are in the frequency intervals of (1.58 to 1.64) d$^{-1}$ and (5.82 to 6.18) d$^{-1}$. They are illustrated with olive green pluses and orange stars, respectively, in Fig.~\ref{fig:lowfreq_pspacing}. Their $\Delta \Pi$ values and errors are reported in Table~\ref{tab:lowfreq_rmode_pspacing}. Their average period spacing values are $\Delta \Pi_{\rm mean_{+}} = 272\pm45$ s and $\Delta \Pi_{\rm mean_{*}} = 148\pm65$ s, respectively.  We suggest that all detected group of modes except the prograde $g$ modes are $r$ modes that are excited because of the fast rotation. We discuss this in details in Sec.~\ref{sec:discussion}. However, there are bunches of frequencies that occur at very low frequencies in the intervals (0.1 and 0.95) d$^{-1}$ and (1.25 to 1.55) d$^{-1}$. These modes are either a linear combination of prograde $g$ modes ($nf_{g_{\rm i}} - mf_{g_{\rm j}}$, $n,m = 1,2$ and $i,j$ as index of the $g$ modes) or prograde $g$ modes and retrograde $r$ modes $f_{8}, f_{24}$ and $f_{35}$ ($f_{g_{\rm i}} - f_{r_{\rm j}}$, $i,j$ representing the $g$ and $r$ modes). Furthermore, there are a few frequencies lower than 0.1 d$^{-1}$. There are also a few significant frequencies in the frequency interval (4.96 to 5.82) d$^{-1}$. These frequencies are either a combination of prograde $g$ modes ($nf_{g_{\rm i}} + mf_{g_{\rm j}}$) or a combination of prograde $g$ and $r$ modes ($af_{g_{\rm i}} + bf_{r_{\rm j}}$ and $a,b = 1,2$). The combinations are listed in Table~\ref{tab:detec_freq}. 
\section{Discussion}\label{sec:discussion}
\subsection{The high-frequency modes}\label{sec:dis_pmodes}
In the region (7 to 23.5) d$^{-1}$, we detected two kinds of multiplets that occur at both sides of the dominant frequency $f_{1}$  (13.97 d$^{-1}$), with frequency spacings $\Delta f_{\rm mean}$ = 1.65 and  0.42 d$^{-1}$. Both regular patterns (which appear in the form of doublets and triplets) are not perfectly symmetrical which could be indicative of second-order rotational effects \citep{Saio1981}. To the first-order approximation, we have \citep{Saio2015}:
\begin{equation}\label{eq:rotation}
\langle P_{\rm rot} \rangle = \frac{1}{\Delta f_{\rm rot}} (1-C_{n,\ell})
\end{equation}
where $C_{n,\ell}$ is the Ledoux constant \citep{Ledoux1951}. For the $p$ modes, the Ledoux constant C$_{\rm n,\ell}$ is zero, and the rotation period is not related to the mode degree $\ell$. Consequently, the rotation period that corresponds to the 1.65 d$^{-1}$ and 0.42 d$^{-1}$ splitting equals P$_{\rm rot} = 0.604$ d and P$_{\rm rot} = 2.385$ d, respectively. $f_{200}$ (1.647$\pm$0.0001 d$^{-1}$) might be the rotation frequency of the fast rotating star. The R value in Table~\ref{tab:photom_data} was obtained from blended photometry, however we know that both components have similar atmospheric properties, thus we may consider the value as a reasonable estimate for each component of the system.  If we consider the radius R = 2.197 R$_{\odot}$ as an estimation of the radius for one of the components, we obtain equatorial velocities V$_{\rm eq}\sim 182$ km/s and $\sim$46 km/s and $i_{\rm rot} \sim 63^{\circ}$ and  $\sim 44^{\circ}$ for the fast and slowly-rotating stars, based on 1.65 d$^{-1}$ and 0.42 d$^{-1}$ splittings, respectively. The consistency between the spectroscopic projected rotational velocity and the one estimated from $\Delta f_{\rm mean}$ indicates that both components of the system may have $p$ modes excited. \\
The three identified combinations of $f_{1}$  with either one of the prograde $g$ modes or one of the retrograde $r$ modes (Mode Couplings) suggests that $f_{1}$, the prograde $g$ and the retrograde $r$ modes are excited in the fast-rotating component. $f_{1}$ is the most dominant frequency and a singlet in $p$ mode region. \\
The \'{e}chelle diagram of the detected $p$ modes belonging to regular patterns shows that the ridges including the modes with different regular frequency splittings (0.42 and 1.65 d$^{-1}$) are crossing each other. This fact reveals these modes are originating from different companion stars \citep[\emph{e.g.}][]{Li2018}. \\ 
\subsection{Regular period spacings of the $g$ modes}
Fig.~\ref{fig:lowfreq_pspacing} illustrates that the excited low-frequency modes include both prograde and retrograde modes. According to \cite{Saio2018b} (Sec. \ref{sec:intro}) we may expect the r modes to appear as groups of low-frequency modes at frequencies slightly lower than $m$ times the rotational frequency, $mf_{\rm rot}$, for intermediate to fast-rotating stars. Effectively, for KIC~8975515, we found a close connection between the identified retrograde modes and the regular frequency spacing $\Delta f_{\rm mean_{\blacksquare}}$ =1.654$\pm$0.018 d$^{-1}$ detected in the high-frequency region. The same conclusion holds for the two small series of frequencies with few modes (the stars and pluses in Fig.~\ref{fig:lowfreq_pspacing}). The retrograde modes appear are in distinct period/frequency regions separated by the harmonics with respect to the harmonics of the (large) rotation frequency ($f_{200}$= 1.647$\pm$0.0001 d$^{-1}$). Hence, we associated these modes to $r$ modes and assigned an azimuthal order $m$ to them according to their position with respect to boundaries of the type $\frac{1}{mf_{\rm rot}}$. For instance, the $r$ modes excited at the frequencies lower than 1.65 d$^{-1}$ (from 1.54 to 1.64 d$^{-1}$) have been assigned to $m =1$, and $m=3$ $r$ modes appear up to the period limit $\frac{1}{3f_{\rm rot}}$ from (3.35 to 4.94) d$^{-1}$. However, a few number of modes with much lower period spacing (73$\pm$13 s) than all other detections are not located in an expected region following such interpretation.Thus, we cannot determine their origin and we did not detect any of them to be in regularly split patterns that could be associated to the rotational frequency of either component (or one of the components). \\
As expected from the models \citep[\emph{e.g.}][]{Saio2018b} the odd and even $r$ modes display different structures: \emph{e.g.}, for $r$ modes $m=3$ the period spacing first increases from 71 s to 215 s and then decreases steeply to 69 s (periods from 0.216 to 0.223 d). In the immediately following period interval, the retrograde pattern starts with $\Delta P$ values increasing from 225 up to 491 s. However, no modes were detected beyond this range as evidence of the (expected) steep decrease of $\Delta P$ close to the period limit ($\frac{1}{2f_{\rm rot}}$). Beyond this period limit, the series of prograde $g$ modes ($m =-2$) appear. In conclusion, we showed that an overwhelming majority of the detected low-frequency modes are located in period intervals which are largely delimited by integer multiples of the rotation frequency of the fast-rotating star. This observational fact supports the conclusion that these $r$ and $g$ modes have the same origin as the series of $p$ modes with $\Delta f_{\rm mean} = 1.65$ d$^{-1}$, i.e. they originate from the fast-rotating companion star. 
\section{Summary}\label{sec:con}
We analysed a 4-year long dataset observed in long-cadence mode by the \emph{Kepler} mission in order to study the pulsations of KIC~8975515. This target consists of a fast-rotating and a slowly rotating companion star (i.e. $v\sin i$ of 162$\pm$2 km/s and 32$\pm$1 km/s) of otherwise similar masses and atmospheric properties. They form a SB2 system in a long-period and eccentric orbit. Binarity in combination with fast rotation of one of the companions along with hybrid pulsations makes this object of high asteroseismic interest.\\ 
The detailed pulsation analysis shows that the most dominant modes in the low- and high-frequency regime have about the same power. Concerning the low-frequency region of the frequency spectrum, the significant pulsation frequencies occur in the range 1.58 - 6.18 d$^{-1}$. We detected five regular period spacing patterns in this regime including a series of prograde $g$ modes and four series of $r$ modes ($m$ = 1, 2, 3 and 4), well located in the frequency intervals delimited by the harmonics of the rotational frequency for the fast-rotating star. \\
We detected two types of combination frequencies in the lower-frequency region which are either a combination of prograde $g$ modes or a combination of a $g$ and a $r$ mode. Both types appear at frequencies lower than $m$=1 up to 0.1 d$^{-1}$ and in-between the $m$=3 and $m$=4 $r$ modes. Some combinations also appear in the intermediate range from 6.72 to 8.63 d$^{-1}$. Concerning the high-frequency regime, the most dominant $p$ mode $f_{1}$ (\emph {i.e.} 13.97 d$^{-1}$) appears as a singlet between two groups of regularly split modes (20 modes between 7.24 and 21.17 d$^{-1}$) with $\Delta f_{\rm rot}$ equal to 0.419$\pm$0.020 d$^{-1}$ and 1.656$\pm$0.018 d$^{-1}$. \\
We interpreted both groups of $p$ modes as multiplets of rotationally split $p$ modes since their mean frequency spacings are in agreement with the estimated rotation velocity of each component. In addition, we detected the presence of the low frequency $f_{200} = 1.647\pm0.001$ d$^{-1}$ whose value clearly agrees with the largest mean frequency spacing supporting the interpretation that it represents the rotation frequency of the fast-rotating star. Furthermore, we identified three combinations of $f_{1}$ with either one of the prograde $g$ modes or one of the retrograde $r$ modes, among the $p$ modes. This indicates that $f_{1}$ has the same origin as the prograde $g$ and the retrograde $r$ modes, \emph{i.e.} they all originate from the fast-rotating component. Finally, the \'{e}chelle diagram of all the detected $p$ mode frequencies versus each one of the mean spacings reveals that the modes associated to the different multiplets are located along crossing ridges, thus each multiplet comes from a different companion star. All of these conclusions put together along with the fact that the regions where the $r$ and $g$ modes appear as distinct groups with respect to the harmonics of the detected (fast) rotation frequency ($f_{200}$) show that we can identify the origin of most of the modes detected in KIC~8975515.\\ 
In summary, we propose on the basis of the presented study that the fast-rotating component is a hybrid ($\gamma$ Dor-$\delta$ Sct) pulsator which also shows $r$ modes of azimuthal order $m$ = 1, 2, 3 and 4 naturally excited because of the fast rotation, while the slow-rotating companion is a $\delta$ Scuti pulsator with regularly (rotationally) split $p$ modes.\\
\begin{acknowledgements}
The authors thank the \emph{Kepler} team efforts to provide and present all the observational data and light curves for the public. The authors also appreciate and thank the \emph{KASOC team} for presenting corrected light curves to the community. The authors acknowledge, the enlightening discussions with \emph{Prof. Donald W. Kurtz} from Jeremiah Horrocks Institute, University of Central Lancashire,  \emph{Dr. Giovanni M. Mirouh} from Astrophysics Research Group, Faculty of Engineering and Physical Sciences, University of Surrey and \emph{Prof. Omar Gustavo Benvenuto} from Universidad Nacional de La Plata, Buenos Aires, Argentina. A. S. also thanks \emph{Paul Van Cauteren} from the Humain Observatory (ROB) for his continuous support. We also acknowledge the anonymous \emph{referee} for very useful comments on the paper. We acknowledge the financial support of \emph{ALMA-CONICYT grant number 31170029} for this research. Furthermore, P.J. acknowledges partial financial support of \emph{FONDECYT Iniciaci\'on grant number 11170174}.  
\end{acknowledgements}
\bibliographystyle{aa}
\bibliography{Samadi.bib}
\appendix
\onecolumn
 \setcounter{table}{0}
\renewcommand{\thetable}{A\arabic{table}}
\begin{longtable}{lccccccc}
 \caption{ \label{tab:lowfreq_rmode_pspacing}Detected $r$ modes in the Fourier spectrum of KIC~8975515 which are located at lower frequencies equal, two and three times ($m = 1,2,3$) of the observational rotational frequency $f_{\rm rot} = 1.65$ d$^{-1}$ (Fig.~\ref{fig:pmode_splt} \& Table~\ref{tab:pmode_splt}) associated to the fast rotating star ($v \sin i = 162$ km\/s). \textit{FR} denotes that the mode originality is from the \textit{F}ast \textit{R}otating star. The determined azimuthal order $m$ is based on $f_{\rm r~mode} = m f_{\rm rot}$. The color coding and symbols are similar to Fig.~\ref{fig:lowfreq_pspacing}. $\epsilon_{\rm \Delta \Pi}$ is the deviation value in seconds from $\Delta \Pi_{\rm mean}$ in each series, that is large due to the retrograde nature of period spacing pattern. The variance error $\epsilon_{\rm var}$ for all of $\Delta \Pi$ is of the order of 10$^{-5}$ sec for each $\Delta \Pi$.}\\
\hline\hline
 $f_\mathrm{i}$  & $f\pm\epsilon_{f}$ & A         & $\Pi$   & $\Delta \Pi \pm\epsilon_{\rm \Delta \Pi}$ & SNR            & color & comment\\
	             & d$^{-1}$           & $\mu$mag  & d       & s                                         &                & Fig. \ref{fig:lowfreq_pspacing}&\\
\hline 
\endfirsthead
\caption{continued.}\\
\hline\hline
 $f_\mathrm{i}$  & $f\pm\epsilon_{f}$ & A         & $\Pi$   & $\Delta \Pi \pm\epsilon_{\rm \Delta \Pi}$ & SNR            & color & comment\\
	             & d$^{-1}$           & $\mu$mag  & d       & s                                         &                & Fig. \ref{fig:lowfreq_pspacing}&\\
\hline
\endhead
\hline
\endfoot
 $f_{\mathrm 57}$  &  1.63391$\pm$0.00006  &  13.63  &  0.61203 &             &  15  & \textcolor{Green}{+} &FR, $m=1$\\
 $f_{\mathrm 24}$  &  1.62692$\pm$0.00004  &  29.96  &  0.61466 &  227$\pm$45 &  31  & \textcolor{Green}{+} &FR, $m=1$\\ 
 $f_{\mathrm 135}$ &  1.59456$\pm$0.00009  &  7.14   &  0.62713 &             &  11  & \textcolor{Green}{+} &FR, $m=1$\\
 $f_{\mathrm 112}$ &  1.58529$\pm$0.00008  &  8.57   &  0.63080 &  317$\pm$45 &  12  & \textcolor{Green}{+} &FR, $m=1$\\ [5pt]   
                   &                       &         &$\Delta \Pi_{\rm mean}\pm \sigma_{\Delta}$&  272$\pm$45&      &      \\[5pt]
 $f_{\mathrm 126}$ &  2.14759$\pm$0.00008  &  7.31   &  0.46564  &             & 9    &\textcolor{OliveGreen}{$\blacktriangle$}  &FR, $m=2$\\     
 $f_{\mathrm 101}$ &  2.13337$\pm$0.00007  &  9.10   &  0.46874  & 268$\pm$88  & 10   &\textcolor{OliveGreen}{$\blacktriangle$}  &FR, $m=2$\\  
 $f_{\mathrm 105}$ &  2.12180$\pm$0.00007  &  9.01   &  0.47130  & 221$\pm$40  & 10   &\textcolor{OliveGreen}{$\blacktriangle$}  &FR, $m=2$\\
\hline                                                            
 \end{longtable} 
 
 \setcounter{table}{1}
\renewcommand{\thetable}{A\arabic{table}}
\begin{landscape}
\begin{longtable}{lccccccccccccc}
 \caption{ \label{tab:detec_freq} Detected frequencies with acceptable signal-to-noise ratio (SNR) from the Fourier spectrum of KIC~8975515 (Long Cadence observations, LC). \textit{P}: parent frequencies (Sec.\ref{sec:fourier}), \textit{SR/FR}: detected $g$/$r$/$p$ modes are most probably associated to Slow Rotating/Fast Rotating star. \textit{PS}: detected $g$/$r$ mode is part of a prograde/retrograde Period Spacing pattern. \textit{FS}: detected $g$/$r$/$p$ mode is with (semi-)regularly spaced from the other frequency in the same multiplet. \textit{MC}: mode is excited as the result of coupling of the most dominant detected $p$ mode ($f_{1}$) and a detected $g$/$r$ mode ($f_{1} \pm f_{g,r}$). The superscript "*" denotes the frequencies that are not formed from a \textit{unique} combination of parent frequencies. $m$ is determined based on $f_{\rm r~mode} = m f_{\rm rot}$. $\ell$ is the mode degree and is determined based on observed rotational splitting of a low-/high-frequency mode.}\\
\hline\hline
 $f_{\mathrm i}$ & $f\pm\epsilon_{f}$  & A$\pm\epsilon_{\mathrm A}$ & $\Pi$ & SNR & P & PS & FS & mode & $m$ &FR/SR & comb. & MC & ref.\struutup\\    
                          & d$^{-1}$                  & $\mu $mag                             & d       &        &    &     &      &          &         &          &             & &\\               
\hline 
\endfirsthead
\caption{continued.}\\
\hline\hline
 $f_{\mathrm i}$ & $f\pm\epsilon_{f}$  & A$\pm\epsilon_{\mathrm A}$ & $\Pi$ & SNR & P & PS & FS & mode & $m$ &FR/SR & comb. & MC& ref.\struutup\\    
                          & d$^{-1}$                  & $\mu $mag                             & d       &        &    &     &      &          &         &          &             & &\\        
\hline
\endhead
\hline
\endfoot
  $f_{ 1}$ &  13.97236$\pm$0.00001    &  267.74$\pm$4.62 &  0.07157  & 204 &   &  &   & $p$ & 0?  & FR & & \struutup\\
 $f_{ 2}$ &  2.37418$\pm$0.00001     &  264.99$\pm$5.31 &  0.42120  &  76 & $\checkmark$ & $\checkmark$ &   & $g$ &  -2? & FR & & & Table~\ref{tab:lowfreq_gmode_pspacing} \& Fig.~\ref{fig:lowfreq_pspacing}\\
 $f_{ 3}$ &  2.31022$\pm$0.00001     &  219.12$\pm$3.99 &  0.43286  &  78 & $\checkmark$ & $\checkmark$ &   & $g$ &  -2? & FR & & & Table~\ref{tab:lowfreq_gmode_pspacing} \& Fig.~\ref{fig:lowfreq_pspacing}\\
 $f_{ 4}$ &  2.52345$\pm$0.00001     &  213.96$\pm$4.68 &  0.39628  &  70 & $\checkmark$ & $\checkmark$ &   & $g$ &  -2? & FR & & & Table~\ref{tab:lowfreq_gmode_pspacing} \& Fig.~\ref{fig:lowfreq_pspacing}\\
 $f_{ 5}$ &  2.25761$\pm$0.00001     &  194.15$\pm$3.28 &  0.44295  &  80 & $\checkmark$ & $\checkmark$ &   & $g$ &  -2? & FR & & & Table~\ref{tab:lowfreq_gmode_pspacing} \& Fig.~\ref{fig:lowfreq_pspacing}\\
 $f_{ 6}$ &  16.07633$\pm$0.00002    &  128.14$\pm$5.18 &  0.06220  &  76 &   &   & $\checkmark$ & $p$ &         & FR & && Table~\ref{tab:pmode_splt} \& Fig.~\ref{fig:pmode_splt}\\
 $f_{ 7}$ &  17.73452$\pm$0.00002    &  84.42$\pm$5.62  &  0.05639  &  76 &   &   & $\checkmark$ & $p$ &         & FR & && Table~\ref{tab:pmode_splt} \& Fig.~\ref{fig:pmode_splt}\\
 $f_{ 8}$ &  2.22490$\pm$0.00002     &  70.43$\pm$3.45  &  0.44946  &  30 & $\checkmark$ & $\checkmark$ & & $g$ &         & FR & & &  \\
\hline                                                            
 \end{longtable}  
 \end{landscape}

\end{document}